\documentclass[pre,preprint,superscriptaddress]{revtex4-1}
\usepackage{epsfig,amsmath,amsfonts,amsthm,graphicx}
\usepackage{ulem}
\usepackage{color}
\usepackage[draft]{changes}
\definechangesauthor[name=Arkady, color=blue]{AP}
\def\vp{\varphi}
\def\e{\varepsilon}

\begin{document}

\title{Synchronization of oscillators with hyperbolic chaotic phases}
\author{Arkady Pikovsky}
\affiliation{Institute of Physics and Astronomy, Potsdam University, 14476 Potsdam-Golm, Germany}
\affiliation{Department of Control Theory, Nizhny Novgorod State University, Gagarin Avenue 23, 603950 Nizhny Novgorod, Russia}
\affiliation{National Research University Higher School of Economics, \\
             25/12 Bolshaya Pecherskaya Ulitsa, 603155 Nizhny Novgorod, Russia}

\begin{abstract}
Synchronization in a population
of oscillators with hyperbolic chaotic phases is studied for two
models. One is based on the Kuramoto dynamics of the phase oscillators and
on the Bernoulli map applied to these phases. This system
possesses an Ott-Antonsen invariant manifold, allowing for a derivation of a
map for the evolution of the complex order parameter. Beyond a critical coupling strength,
this model demonstrates bistability synchrony-disorder. Another model
is based on the coupled autonomous oscillators with hyperbolic chaotic strange
attractors of Smale-Williams type. Here a disordered asynchronous state at small coupling strengths, and
a completely synchronous state at large couplings are observed. Intermediate
regimes are characterized by different levels of complexity of the 
global order parameter dynamics.
\end{abstract}
\maketitle
\section{Introduction}
Synchronization of chaotic oscillators has many aspects~\cite{Pikovsky-Rosenblum-Kurths-01},
one generally distinguishes complete, generalized, and phase synchronization.
The latter property is related to chaotic oscillators with well-defined phases.
Many chaotic oscillators, like the R\"ossler system, possess chaotic amplitudes, while
the phase in such systems is not chaotic and corresponds to a zero Lyapunov exponent.
The dynamics of the phase is of diffusion type, and correspondingly the phase synchronization 
phenomena for such oscillators are close to those for periodic oscillators with
a certain level of noise in the phase dynamics.

In a seminal paper~\cite{Kuznetsov-05} S. P. Kuznetsov constructed a physical
model of an oscillator with a \textit{chaotic phase}. In this construction,
the process has amplitude modulation, and at each period of
modulation the phase experience a doubling map. The overall attractor is hyperbolic
and
belongs to a Smale-Williams solenoid class. In a series of subsequent publications,
summarized in the book~\cite{Kuznetsov-12}, S. P. Kuznetsov and co-authors provided many
examples of systems with hyperbolic phase chaos.

In this paper we study synchronization properties of the oscillators with
chaotic phases. First, we construct a rather abstract model, where phase chaos and 
synchronizing interactions are separated in time (Section \ref{sec:kbm}). Namely, the process
consists of two epochs: in one epoch phase oscillators interact according
to the Kuramoto global coupling scheme, and in another epoch the phases undergo
a chaotic Bernoulli map. This model demonstrates, for certain values of parameters,
a bistability between a desynchronized and a synchronized states. In Section \ref{sec:kpm}
we consider coupled autonomous oscillators with chaotic phases, constructed
by S. P. Kuznetsov and the author in~\cite{Kuznetsov-Pikovsky-07}. This
system demonstrates a rather reach behavior with asynchronous, completely 
synchronous, and complex partially synchronous states.

\section{Kuramoto-Bernoulli model}
\label{sec:kbm}
In this section we construct a model of interacting phase oscillators,
which combines features of the Kuramoto model~\cite{Acebron-etal-05} 
(global attractive coupling of the phases) with the
hyperbolic chaotic dynamics of the phases described by a Bernoulli map.
\subsection{Kuramoto ensemble and OA evolution}
Consider $N$ phase oscillators $\vp_k$ interacting via Kuramoto mean-field coupling
\begin{equation}
\dot\vp_k=\mu R\sin(\Theta-\vp_k),\quad Z=R e^{i\Theta}=\frac{1}{N}\sum_k e^{i\vp_k}\;.
\label{eq:km}
\end{equation}
Here $Z$ is the complex mean field, and $\mu$ is the coupling constant.
Quantity $R$ is called Kuramoto order, it characterizes asynchronous ($R=1$)
and synchronous ($R=1$)  regimes.
We assume that all the oscillators have the same frequency, and write equations
in the reference frame where this frequency vanishes, so it does not enter 
in~\eqref{eq:km}. For $\mu>0$ the coupling is attractive,
and in this situation all the oscillators eventually synchronize: $R\to 1$,
and a state where $\vp_1=\vp_2=\ldots \vp_N$ establishes.

Synchronization transition is monotonous (in fact, there exists 
a Lyapunov function that governs it), but it can be generally hardly 
expressed analytically. 
An analytic solution is, however, possible,
if the Ott-Antonsen (OA) ansatz~\cite{Ott-Antonsen-08}, which
applies to the thermodynamic limit $N\to\infty$, is peformed.
In the OA ansatz it is assumed that
the distribution of the phases is a wrapped Cauchy
distribution, and the complex circular moments
\begin{equation}
Z_k=\langle e^{i k\vp}\rangle
\label{eq:op}
\end{equation}
can all be expressed via the complex mean field $Z_k=Z^k$.
Then the equation for the order
parameter reads~\cite{Ott-Antonsen-08}
\[
\dot R=\frac{\mu}{2}R(1-R^2)\;.
\]
Evolution of the complex mean field during a time interval $T$ is
\begin{equation}
R(T)=\frac{R(0)}{\sqrt{R^2(0)+(1-R^2(0))\exp[-\mu T]}}
\label{eq:kd}
\end{equation}
One can see that the only parameter in this transformation is $\gamma=\exp(-\mu T)$.
Evidently, $R\to 1$ as $n\to infty$, and the rate of this convergence
is larger for smaller $\gamma$.
\subsection{Bernoulli map of phases}
Consider a Bernoulli map acting on the phases 
\begin{equation}
\vp(n+1)= K \vp(n)\;,
\label{eq:bm}
\end{equation}
 with an integer parameter $K$.
For an ensemble of Bernoulli maps~\eqref{eq:bm}, it is easy to express the evolution
of the probability density of phases through the complex circular moments \eqref{eq:op}:
\[
Z_m(n+1)=Z_{Km}(n)\;.
\]
One can see that the OA ansatz is invariant under Bernoilli maps.
Indeed, if $Z_m(n)=Z^m$, then $Z_m(n+1)=Z^{Km}=(Z^K)^m$. Thus,
the evolution of the complex mean field under the Bernoulli map is
\begin{equation}
Z(n+1)=Z^K (n)\;.
\label{eq:mfbm}
\end{equation}

\subsection{Kuramoto ensemble and Bernoilli map}
We construct a Kuramoto-Bernoulli (KB) model as a sequence of applications
of the Kuramoto dynamics \eqref{eq:kd} and of the Bernoulli dynamics
\eqref{eq:mfbm}. Application of the expressions \eqref{eq:kd}, \eqref{eq:mfbm}
leads to the following map for the order parameter
\[
R(n+1)=\frac{R^K(n)}{(R^2(n)+(1-R^2(n))\gamma)^{K/2}}
\]
This map has always a stable asynchronous 
fixed point $R_{as}=0$, and a synchronous fixed point $R_{s}=1$.
The fixed point $R=1$ is stable for
\begin{equation}
\gamma<\frac{1}{K}\;,
\label{eq:st}
\end{equation}
in this case also an unstable partially synchronous fixed point with $0<R_{ps}<1$ exists,
so there is a bistability asynchrony-synchrony.

The threshold for synchrony stability \eqref{eq:st} is valid not only in the OA approximation, but generally.
Indeed, close to the synchronous state the deviations of the phases satisfy, in the Kuramoto stage, 
the linear equation
\[
\frac{d}{dt}\delta\vp=-\mu\delta\vp
\]
so that combined map for the linear deviations is
\[
\delta\vp(n+1)=K\gamma\delta\vp(n)
\]
from which \eqref{eq:st} follows.

\begin{figure}
\centering
\includegraphics[width=\columnwidth]{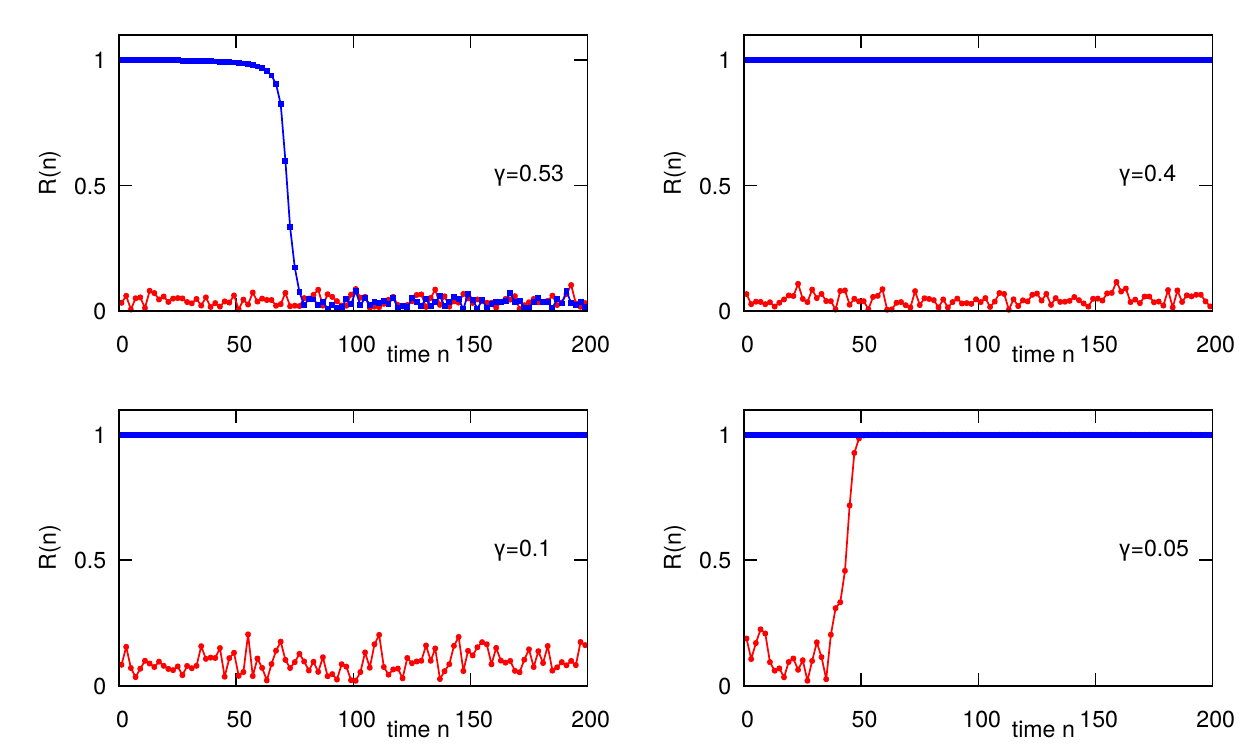}
\caption{Illustration of the dynamics in a Kuramoto-Bernoulli model
with $K=2$ and different $\gamma$, ensemble size $N=1000$.  
Red curves: disordered initial state; blue curves:
ordered initial state (distribution of phases in an interval $(0,0.1)$).
}
\label{fig:id}
\end{figure}

We illustrate the dynamics of the KB model in Fig.~\ref{fig:id}. There we show
the evolution of the oder parameter $R$ for different values of parameter $\gamma$ and 
different initial states. The fully synchronous state is absorbing 
(exactly the same phases remain the same) for all system sizes $N$,
while
there are finite-size fluctuations around the disordered state. 
For small $\gamma$,
one observes a finite-size induced transition to the synchronous state. 

\section{Globally coupled Kuznetsov-Pikovsky (KP) oscillators}
\label{sec:kpm}
Here we study globally coupled chaotic phase oscillators introduced by S. P. Kuznetsov and the author
in
Ref.~\cite{Kuznetsov-Pikovsky-07}.

\subsection{One KP oscillator}
\begin{figure}
\centering
(a)\includegraphics[width=0.45\columnwidth]{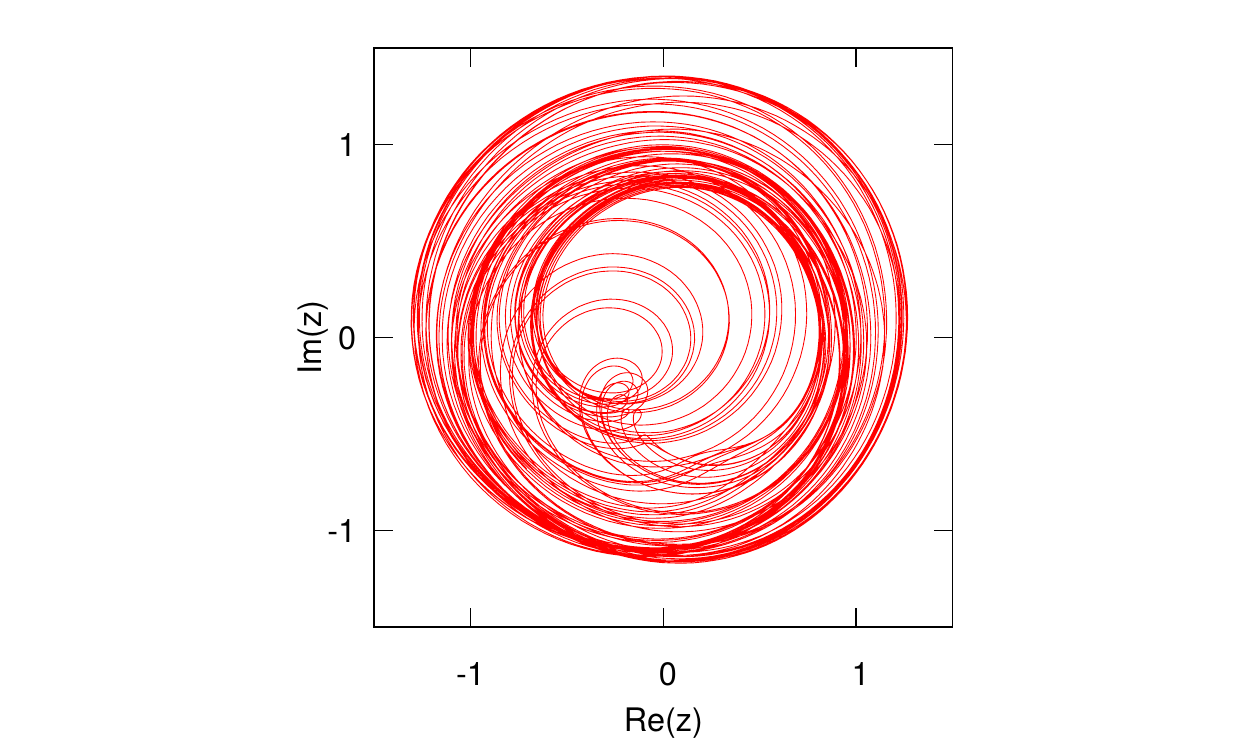}\hfill
(b)\includegraphics[width=0.43\columnwidth]{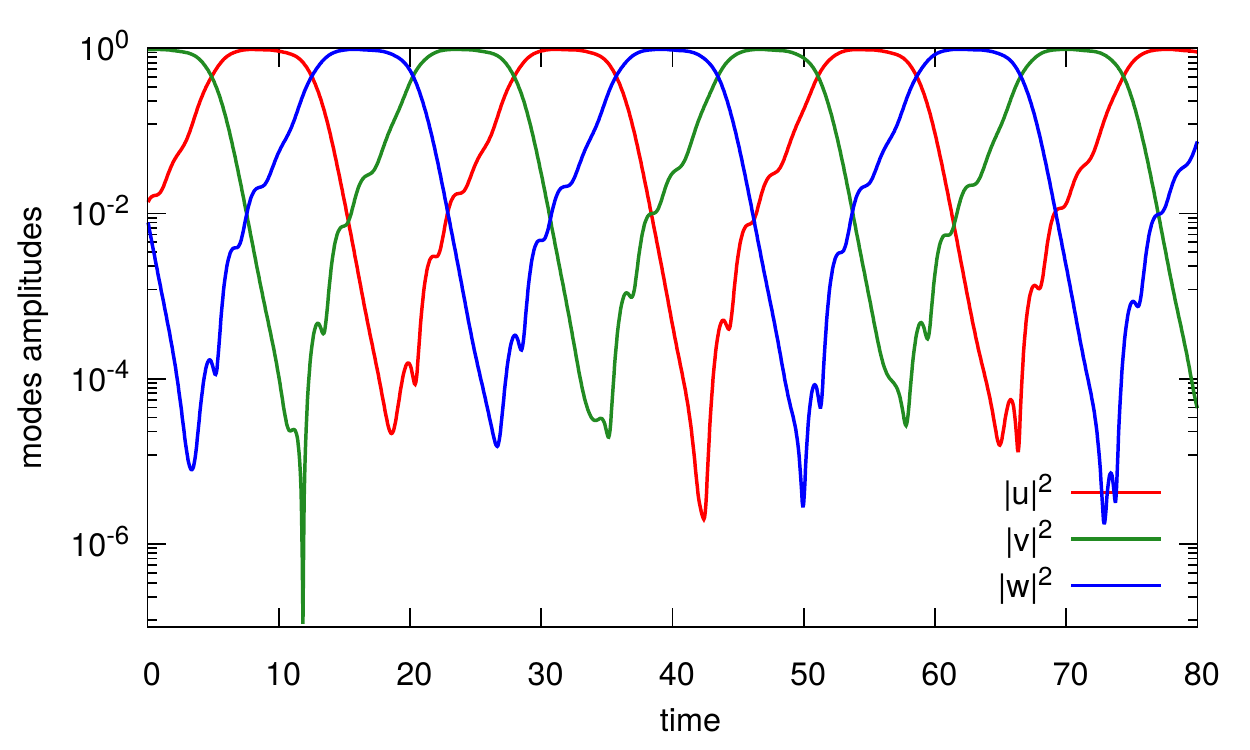}
\caption{Illustration of the dynamics in a KP model.  
Panel (a): phase portrait of the observable $z=u+v+w$ (complex amplitude of oscillations). 
Panel (b): amplitudes of the modes.}
\label{fig:oo1}
\end{figure}

\begin{figure}
\centering
\includegraphics[width=0.4\columnwidth]{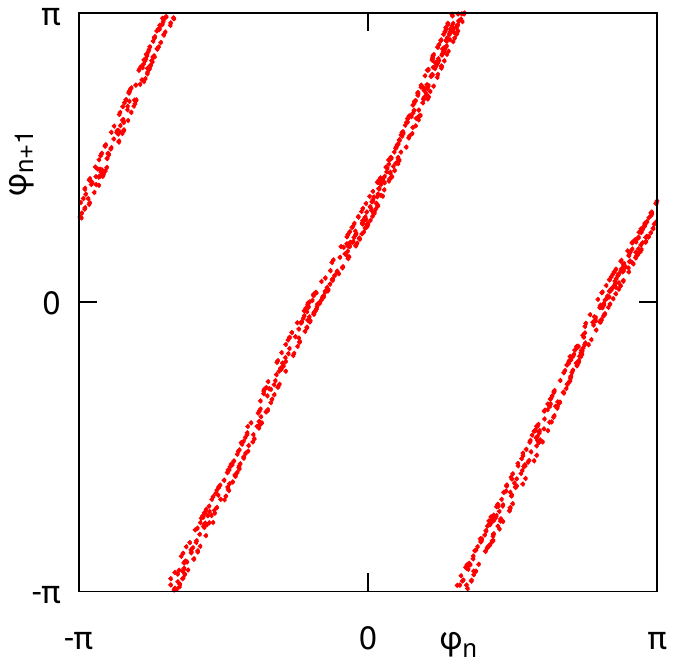}
\caption{Illustration of the phase transformation in model ~\eqref{eq:kp}.
The consecutive phases $\text{arg}(u)\to \text{arg}(v)\to\text{arg}(w)$ are depicted.
The transformation $\text{arg}(u)\to\text{arg}(u)$ would be $\vp\to 2^3\vp$.}
\label{fig:oo2}
\end{figure}

An individual KP oscillator consists of three modes described by their
complex amplitudes $u,v,w$. The equation of one unit are
\begin{equation}
\begin{aligned}
\dot u&=-i u +(1-|u|^2-\frac{1}{2}|v|^2-2|w|^2)u +\e\text{Im}(v^2)\;,\\
\dot v&=-i v +(1-|v|^2-\frac{1}{2}|w|^2-2|u|^2)v +\e\text{Im}(w^2)\;,\\
\dot w&=-i w +(1-|w|^2-\frac{1}{2}|u|^2-2|v|^2)u +\e\text{Im}(u^2)\;.
\end{aligned}
\label{eq:kp}
\end{equation}
Below we fix the internal coupling parameter $\e=0.075$. 
For $\e=0$, system \eqref{eq:kp} has a stable homoclinic cycle,
where the modes are excited consequentially $w\to v\to u \to w\to\ldots$,
with increasing periods of the cycle. The effect of coupling $\e>0$ is twofold:
first, the cycle period is limited from above (see Fig.~\ref{fig:oo1}(b)),
and second, at each stage where a mode amplitude passes close to zero,
its phase attains the doubled value of the exciting mode. The latter property
is described in Ref.~\cite{Kuznetsov-Pikovsky-07} in details; here
we illustrate it with figure \ref{fig:oo2}. Thus, the KP oscillator \eqref{eq:kp}
has a chaotic phase obeying a Bernoulli map.

\subsection{Globally coupled KP oscillators}
Here we introduce a global coupling of $N$ oscillators (numbered by index $k$), 
such that complete synchrony is possible:
\begin{equation}
\begin{aligned}
\dot u_k&=-i u_k +(1-|u_k|^2-\frac{1}{2}|v_k|^2-2|w_k|^2)u +\e\text{Im}(v_k^2)
+\mu|u_k|^2(U-u_k),\\
\dot v_k&=-i v_k +(1-|v_k|^2-\frac{1}{2}|w_k|^2-2|u_k|^2)v +\e\text{Im}(w_k^2)
+\mu|v_k|^2(V-v_k),\\
\dot w_k&=-i w_k +(1-|w_k|^2-\frac{1}{2}|u_k|^2-2|v_k|^2)u +\e\text{Im}(u_k^2)
+\mu| w_k|^2(W-w_k),\\
&U=\frac{1}{N}\sum_k u_k,\quad V=\frac{1}{N}\sum_k v_k,\quad 
W=\frac{1}{N}\sum_k w_k\;.
\end{aligned}
\label{eq:gckp}
\end{equation}
The coupling term is proportional to the parameter $\mu$,
it contains three complex mean fields $U,V,W$, corresponding to three
modes of each oscillator.

\begin{figure}
\centering
\includegraphics[width=\columnwidth]{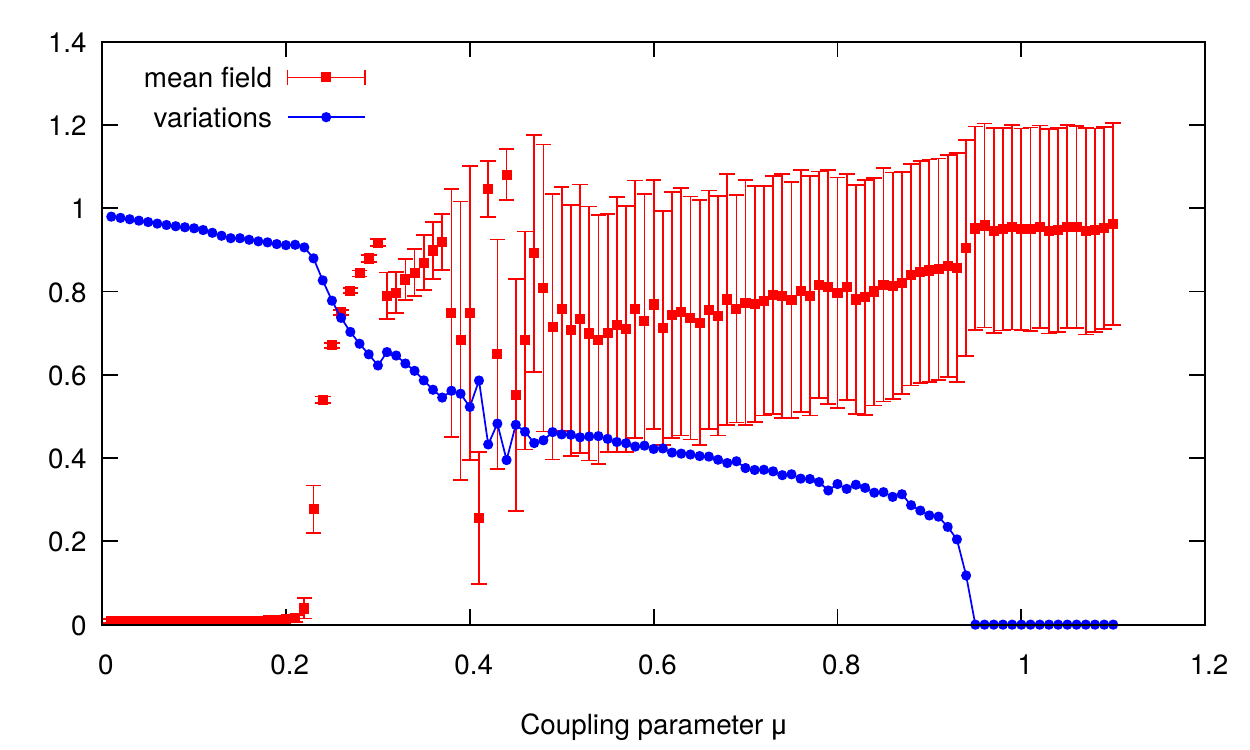}
\caption{Bifurcation diagram of model \eqref{eq:gckp} for $N=10^4$.}
\label{fig:ebd}
\end{figure}

Figure~\ref{fig:ebd} intends to
characterize the dynamical regimes in the system, in dependence on the
coupling parameter $\mu$.
Here two quantities have been calculated. First, I present the dynamics
of the global complex mean field
\[
Z(t)=U(t)+V(t)+W(t)\;.
\]
I calculated the time average $\langle |Z|\rangle_t$ and its
fluctuations $\langle (|Z|-\langle |Z|\rangle_t)^2\rangle_t$, these quantities are shown
in Fig.~\ref{fig:ebd} with red (fluctuations as error bars). Additionally, 
for each moment of time, 
I calculated the spread in the ensemble
\[
D(t)=\frac{1}{N}\sum_k \left[|u_k(t)-U(t)|^2+|v_k(t)-V(t)|^2+|w_k(t)-W(t)|^2\right]
\]
and then average this quantity over time. This quantity is shown with blue.

\begin{figure}
\centering
(a)\includegraphics[width=0.7\columnwidth,clip]{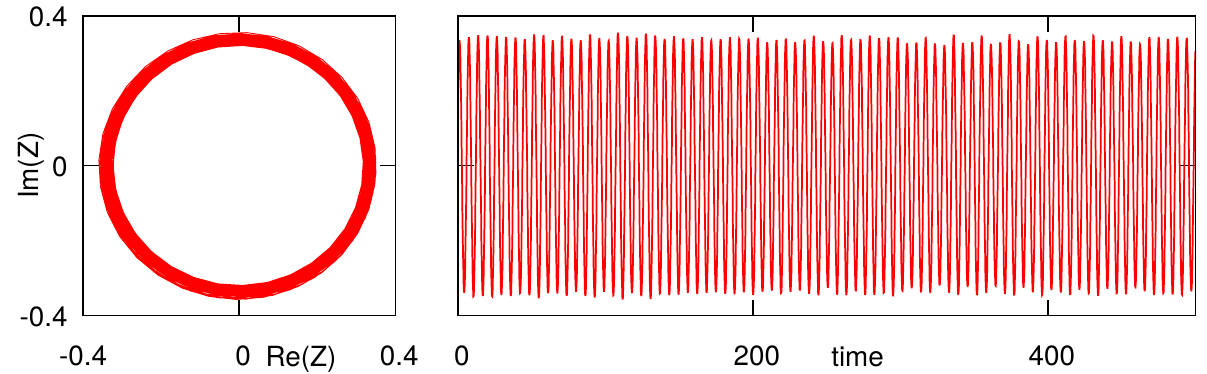}
(b)\includegraphics[width=0.7\columnwidth,clip]{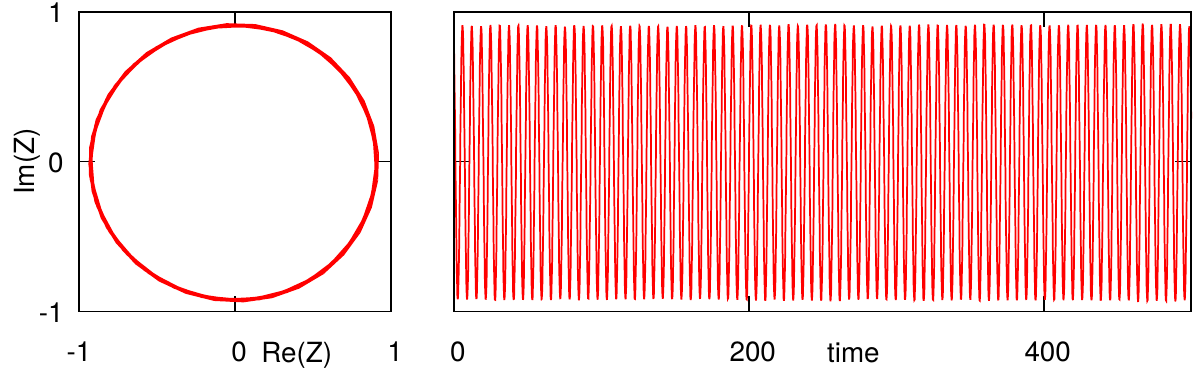}
(c)\includegraphics[width=0.7\columnwidth,clip]{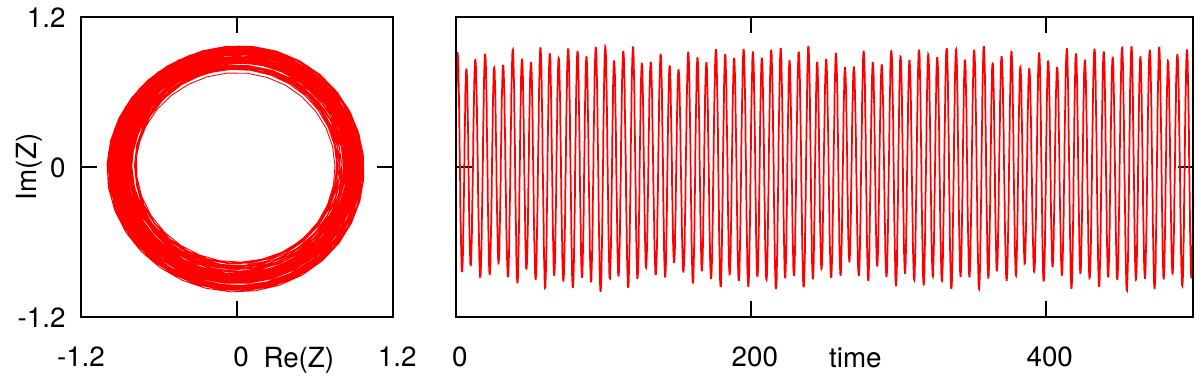}
\caption{The dynamics of the global complex order parameter
for $\mu=0.23$ (panel (a)), $\mu=0.3$ (panel (b)), and $\mu=0.35$ (panel (c)).}
\label{fig:f1}
\end{figure}

\begin{figure}
\centering
(a)\includegraphics[width=0.7\columnwidth,clip]{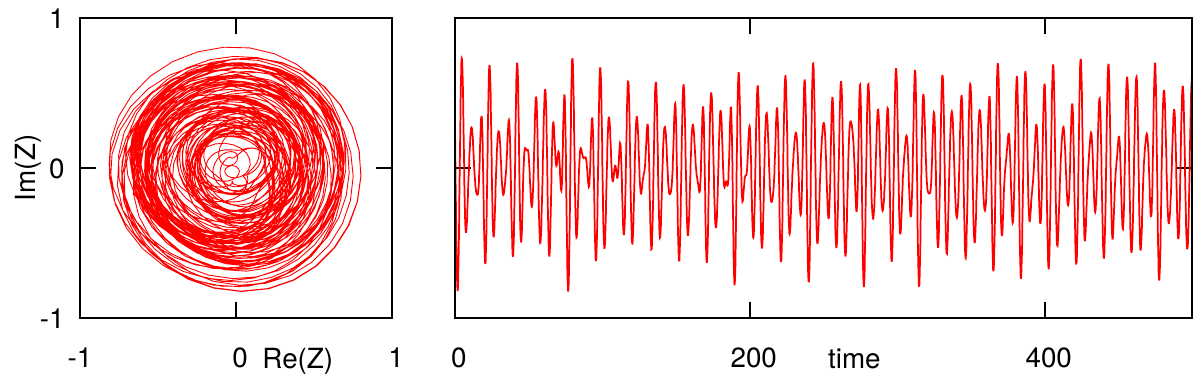}
(b)\includegraphics[width=0.7\columnwidth,clip]{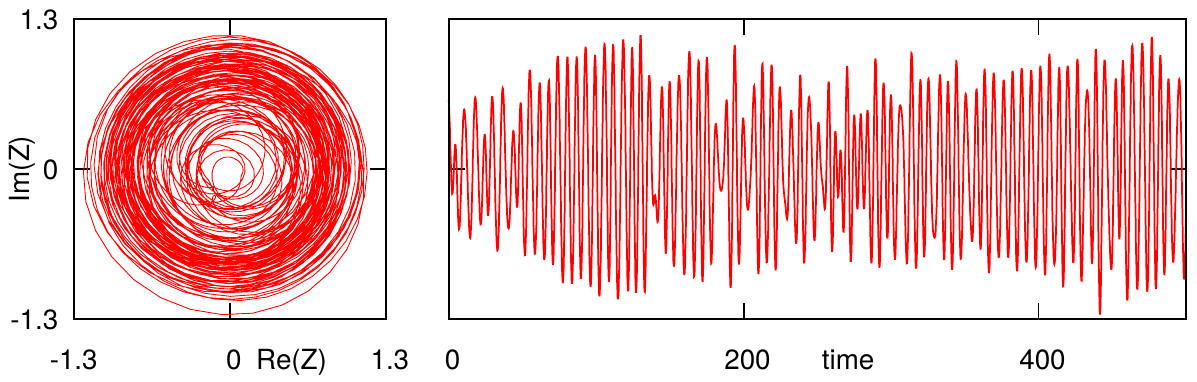}
(c)\includegraphics[width=0.7\columnwidth,clip]{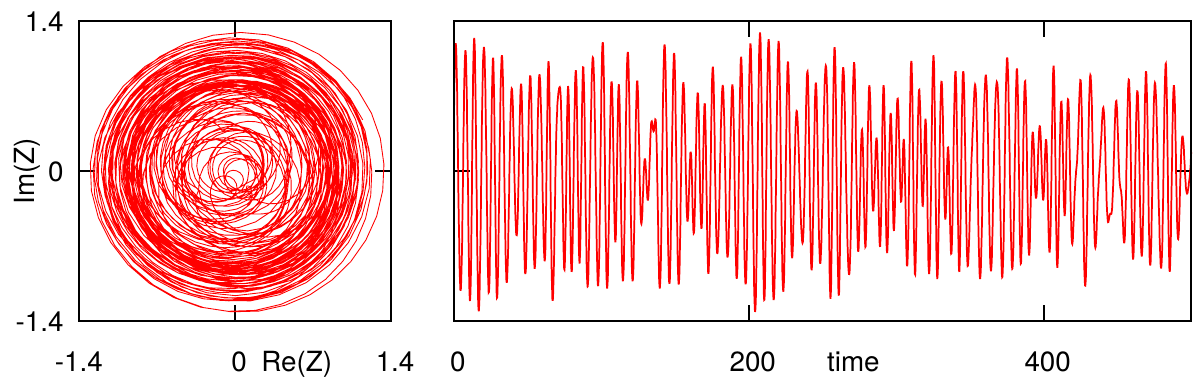}
\caption{The dynamics of the global complex order parameter
for $\mu=0.45$ (panel (a)), $\mu=0.6$ (panel (b)), and $\mu=0.8$ (panel (c)).}
\label{fig:f2}
\end{figure}

Below I  describe different states on the bifurcation diagram.
\begin{enumerate}
\item \textbf{Complete synchronization.} This regime is observed for $\mu>0.95$. Here
$u_k=u_j,v_k=v_j,w_k=w_j$ for all $k,j$. In this state $D=0$.
\item \textbf{Asynchronous state.} This regime is observed for $\mu \lesssim 0.22$.
Here the mean field vanishes, and one has effectively a set of non-interacting oscillators.
\item \textbf{Periodic mean field.} This regime is observed in the range
$0.22 \lesssim \mu\lesssim 0.31$. Here the complex mean field $Z(t)$ is nearly periodic.
We illustrate this in Fig.~\ref{fig:f1}(a,b). There are visible fluctuations for $\mu=0.23$,
but
for $\mu=0.3$ periodicity is nearly perfect. The transition to a periodic mean field
at $\mu\approx 0.22$ is very much similar to one described in 
Ref.~\cite{Pikovsky-Rosenblum-Kurths-96}.
\item \textbf{Weakly irregular mean field} This state is illustrated in  Fig.~\ref{fig:f1}(c).
At $\mu=0.35$ the mean field is close to periodic one, but has a seemingly nearly quasiperiodic modulation.
\item \textbf{Irregular mean field} This state is observed for $0.4\lesssim \mu \lesssim 0.95$,
we illustrate it  in  Fig.~\ref{fig:f2} (a-c). Fluctuations of the mean field are essential, eventually for large $\mu$
they become close to the fluctuations of the field $z(t)$ in one chaotic oscillator.
\end{enumerate}

%

\section{Discussion}
In this paper I studied effects of coupling on oscillators with hyperbolic chaotic dynamics
of the phases. In the simplest, rather artificial Kuramoto-Bernoulli model,
an exact mapping for the order parameter has been derived in the Ott-Antonsen 
approximation in the thermodynamic limit. 
The dynamics here, beyond a certain level of coupling, is bistable:
synchronous and asynchronous states coexist. In relatively small ensembles, for
strong enough coupling, only synchronous states survives
as it is a truly absorbing one. A more realistic model of coupled autonomous continuous-time
oscillators with hyperbolic dynamics of the phases demonstrated much more rich dynamics.
Together with a fully asynchronous state at small coupling strengths, and
a completely synchronous at strong coupling strengths, it demonstrates different states with partial synchrony.
Close to the asynchronous state, the mean field is nearly periodic; and with increase of
coupling strength it becomes irregular through presumably a quasi-periodic state. Detailed study of partially 
synchronous states in this model will be a subject of a separate study.

\begin{acknowledgments}
A. P.
acknowledges support by the Russian Science Foundation (studies
of Section~\ref{sec:kpm}, grant Nr. 17-12-01534) 
and by DFG (grant PI 220/21-1). Numerical experiments in Sec.~\ref{sec:kbm} 
were supported by the Laboratory of Dynamical Systems and Applications NRU HSE, 
of the Russian Ministry of Science and Higher Education (Grant No. 075-15-2019-1931).
\end{acknowledgments}

\def\cprime{$'$}

\end{document}